\documentstyle[12pt]{article}
\newlength{\extraspace}
\newlength{\extraspaces}
\catcode`\@=11
\def\numberbysection{\@addtoreset{equation}{section}
\def\theequation{\arabic{section}.\arabic{equation}}}
\newcommand{\be}{\begin{equation}
\addtolength{\abovedisplayskip}{\extraspaces}
\addtolength{\belowdisplayskip}{\extraspaces}
\addtolength{\abovedisplayshortskip}{\extraspace}
\addtolength{\belowdisplayshortskip}{\extraspace}}
\newcommand{\ee}{\end{equation}}
\newcommand{\ba}{\begin{eqnarray}
\addtolength{\abovedisplayskip}{\extraspaces}
\addtolength{\belowdisplayskip}{\extraspaces}
\addtolength{\abovedisplayshortskip}{\extraspace}
\addtolength{\belowdisplayshortskip}{\extraspace}}
\newcommand{\ea}{\end{eqnarray}}
\newcommand{\nonu}{\nonumber \\[.5mm]}
\newcommand{\A}{&\!\!\!}

\begin{document}
\addtolength{\baselineskip}{.7mm}
\thispagestyle{empty}
\begin{flushright}
SIT--LP--01/06 \\
STUPP--01--162 \\
{\tt hep-th/0110102} \\ 
October, 2001
\end{flushright}
\vspace{7mm}
\begin{center}
{\Large{\bf On Linearization of $N=1$ Nonlinear \\[2mm]
Supersymmetry 
}} \\[20mm] 
{\sc Kazunari Shima}${}^{\rm a}$\footnote{
\tt e-mail: shima@sit.ac.jp}, \ 
{\sc Yoshiaki Tanii}${}^{\rm b}$\footnote{
\tt e-mail: tanii@post.saitama-u.ac.jp} \ 
and \ 
{\sc Motomu Tsuda}${}^{\rm a}$\footnote{
\tt e-mail: tsuda@sit.ac.jp} 
\\[5mm]
${}^{\rm a}${\it Laboratory of Physics, 
Saitama Institute of Technology \\
Okabe-machi, Saitama 369-0293, Japan} \\[3mm]
${}^{\rm b}${\it Physics Department, Faculty of Science, 
Saitama University \\
Saitama 338-8570, Japan} \\[20mm]
{\bf Abstract}\\[7mm]
{\parbox{13cm}{\hspace{5mm}
The $N=1$ Volkov-Akulov model of nonlinear supersymmetry is  
explicitly related to a vector supermultiplet model with a 
Fayet-Iliopoulos $D$ term of linear supersymmetry. 
The physical significance of the results is discussed briefly. 
}}
\end{center}
\vfill
\newpage
\setcounter{section}{0}
\setcounter{equation}{0}
%
%
%
Spontaneous breakdown of supersymmetry (SUSY) produces 
inevitably Nambu-Goldstone (N-G) fermions \cite{SS}, 
as demonstrated in the Fayet-Iliopoulos model \cite{FI} 
and the O'Raifeartaigh model \cite{O}. 
Dynamics of N-G fermions is 
described by the Volkov-Akulov action \cite{VA}. 
When N-G fermions are coupled to supergravity \cite{FNFDZ} 
under a local SUSY invariant way, they are converted to 
the longitudinal components of spin 3/2 fields by the super Higgs 
mechanism \cite{DZ} as demonstrated for the V-A model. 
\par
This may be always the case if we adhere to  the coset space 
G/H interpretation of the  nonlinear realization 
of SUSY and to the assumption of the existence of the invariant 
action under the initial larger symmetry group G with local 
SUSY. Most of the SUSY unified theories adopt  this mechanism and 
N-G fermions disappear at low energy, which gives an 
explanation of the absence of free (bare) N-G fermions in nature. 
\par
However if we consider seriously the distinguished character of 
SUSY \cite{WZ1}, i.e., SUSY and its spontaneous breakdown are 
profoundly connected to the noncompact spacetime (Poincar\'e) 
symmetry, it may be worthwhile regarding the V-A model 
as a nonlinear realization of SUSY originated not necessarily 
from specific Lagrangian models of G and G/H expressed by field 
operators but from a spontaneous breakdown of the higher symmetry 
of spacetime by itself in terms of the geometrical arguments. 
\par
In ref.\ \cite{KS1} one of the authors has 
proposed the {\it superon-graviton model} (SGM) as an attempt 
along this idea. The fundamental action of the SGM is an 
Einstein-Hilbert action analogue of general relativity, 
which is obtained by the geometrical arguments of the local 
GL(4,{\bf R}) invariance of the SGM spacetime, 
where there exist fermionic degrees of freedom (N-G fermions) 
at every four-dimensional curved spacetime point. 
It consists of the Einstein-Hilbert action, the V-A action 
with a global SO(10) and their interactions and is invariant under 
a new SUSY \cite{ST1}. All observed (low energy) elementary 
particles except graviton are 
regarded as (composite) eigenstates of the {\it linear} 
representation of the SO(10) super Poincar\'e algebra composed of 
fundamental objects {\it superons} (N-G fermions) with spin 1/2 
\cite{KS2}. For deriving the low energy physical contents of 
the SGM action it is often useful to linearize  such a highly 
nonlinear theory and obtain a low energy effective theory. 
Toward the linearization of the SGM we investigate the 
linearization of the V-A model in detail. 
\par
The linearization of the V-A model was investigated by many 
authors \cite{IK,R,UZ,SW,NISHINO}. 
Ivanov and Kapustnikov \cite{IK} have established the general 
relations between linear and nonlinear 
realizations of global SUSY.  
In ref.\ \cite{R} Ro\v{c}ek constructed irreducible and SUSY 
invariant constraints on a scalar supermultiplet in terms of 
the N-G field and  showed explicitly that the V-A model of 
nonlinear SUSY was related to a scalar supermultiplet of the 
linear SUSY of Wess and Zumino \cite{WZ1}.
In ref.\ \cite{IK} a relationship between the V-A model and 
a vector supermultiplet is studied in terms of a constrained 
gauge superfield in the context of the coupling of the V-A 
action to the gauge multiplet action with the Fayet-Iliopoulos 
$D$ term of linear SUSY. 
Although the relation between the action of linear SUSY and 
the V-A action is established as expected from the viewpoint 
that they are equally responsible to spontaneous SUSY breaking. 
The explicit representation of all component fields of the 
vector superfield in terms of the N-G fermion, which is 
crucial for the SGM scenario, is remained to be studied. 
\par
In this letter  we construct the complete form of the SUSY 
invariant constraints and show  explicitly that the V-A model is 
related to the total action of a U(1) gauge supermultiplet 
\cite{WZ2} of the linear SUSY with the Fayet-Iliopoulos $D$ 
term indicating a spontaneous SUSY breaking. 
We find that a U(1) gauge field can be constructed explicitly 
from the N-G fermion fields although it is an axial vector. 
\par
%
%
An $N=1$ U(1) gauge supermultiplet is given by a real superfield 
\ba
V(x, \theta, \bar\theta) 
\A = \A C + i \theta \chi - i \bar\theta \bar\chi 
+ {1 \over 2} i \theta^2 (M+iN) 
- {1 \over 2} i \bar\theta^2 (M-iN) \nonu
\A\A - \theta \sigma^m \bar\theta v_m 
+ i \theta^2 \bar\theta \left( \bar\lambda 
+ {1 \over 2} i \bar\sigma^m \partial_m \chi \right) 
- i \bar\theta^2 \theta \left( \lambda 
+ {1 \over 2} i \sigma^m \partial_m \bar\chi \right) \nonu
\A\A + {1 \over 2} \theta^2 \bar\theta^2 
\left( D + {1 \over 2} \Box C \right), 
\ea
where $C(x)$, $M(x)$, $N(x)$, $D(x)$ are real scalar fields, 
$\chi_\alpha(x)$, $\lambda_\alpha(x)$ and 
$\bar\chi_{\dot\alpha}(x)$, $\bar\lambda_{\dot\alpha}(x)$ are 
Weyl spinors and their complex conjugates, and $v_m(x)$ is 
a real vector field. 
We use the two-component spinor notation in ref.\ \cite{WB}. 
Spacetime vector indices are denoted by $m, n, \cdots 
= 0, 1, 2, 3$, and spinor indices by $\alpha, \beta, \cdots 
= 1, 2$ and $\dot\alpha, \dot\beta, \cdots = 1, 2$. 
For more details of the notations see ref.\ \cite{WB}. 
Only $\lambda$, $\bar\lambda$, $D$ and 
$v_{mn} = \partial_m v_n - \partial_n v_m$ are gauge invariant. 
Other component fields can be set to zero by a gauge 
transformation in the superspace. 
The supertransformation of $V$ with transformation parameters 
$\epsilon_\alpha$, $\bar\epsilon_{\dot\alpha}$ is given by 
\be
\delta V = \left( \epsilon Q + \bar\epsilon \bar Q \right) V, 
\label{supertrans}
\ee
where 
\be
Q_\alpha = {\partial \over \partial \theta^\alpha} 
- i (\sigma^m {\bar\theta})_\alpha \partial_m, \qquad
\bar Q_{\dot\alpha} 
= -{\partial \over \partial {\bar\theta}^{\dot\alpha}} 
+ i ( \theta \sigma^m)_{\dot\alpha} \partial_m. 
\ee
\par
We introduce an N-G fermion field $\zeta_\alpha(x)$ and 
its complex conjugate $\bar\zeta_{\dot\alpha}(x)$. 
Their supertransformations are 
\ba
\delta \zeta \A = \A {1 \over \kappa} \epsilon - i \kappa \left( 
\zeta \sigma^m \bar\epsilon - \epsilon \sigma^m \bar\zeta \right) 
\partial_m \zeta, \nonu
\delta \bar\zeta \A = \A {1 \over \kappa} \bar\epsilon - i \kappa 
\left( \zeta \sigma^m \bar\epsilon - \epsilon \sigma^m \bar\zeta 
\right) \partial_m \bar\zeta, 
\label{zetasupertrans}
\ea
where $\kappa$ is a constant whose dimension is $({\rm mass})^{-2}$. 
{}Following refs.\ \cite{IK}, \cite{UZ} we define the superfield 
$\tilde V(x, \theta, \bar\theta)$ by 
\be
\tilde V(x, \theta, \bar\theta) = V(x', \theta', \bar\theta'), 
\label{tildev}
\ee
where 
\ba
x'^{\,m} \A = \A x^m + i \kappa \left( 
\zeta(x) \sigma^m \bar\theta 
- \theta \sigma^m \bar\zeta(x) \right), \nonu
\theta' \A = \A \theta - \kappa \zeta(x), \qquad
\bar\theta' = \bar\theta - \kappa \bar\zeta(x). 
\label{cov}
\ea
$\tilde V$ may be expanded in component fields as 
\ba
\tilde V(x, \theta, \bar\theta) 
\A = \A \tilde C + i \theta \tilde\chi 
- i \bar\theta \bar{\tilde\chi} 
+ {1 \over 2} i \theta^2 (\tilde M+i\tilde N) 
- {1 \over 2} i \bar\theta^2 (\tilde M-i\tilde N) \nonu
\A\A - \theta \sigma^m \bar\theta \tilde v_m 
+ i \theta^2 \bar\theta \left( \bar{\tilde\lambda} 
+ {1 \over 2} i \bar\sigma^m \partial_m \tilde\chi \right) 
- i \bar\theta^2 \theta \left( \tilde\lambda 
+ {1 \over 2} i \sigma^m \partial_m \bar{\tilde\chi} \right) \nonu
\A\A + {1 \over 2} \theta^2 \bar\theta^2 
\left( \tilde D + {1 \over 2} \Box \tilde C \right), 
\label{tildeexp}
\ea
where $\tilde C, \tilde\chi, \bar{\tilde\chi}, \cdots$ can be 
expressed by $C, \chi, \bar\chi, \cdots$ and $\zeta$, $\bar\zeta$ 
by using the relation (\ref{tildev}). From eqs.\ (\ref{supertrans}), 
(\ref{zetasupertrans}) it can be shown that supertransformations 
of these component fields 
$\tilde\phi_i(x) = (\tilde C, \tilde\chi, \bar{\tilde\chi}, 
\cdots)$ have a form 
\be
\delta \tilde\phi_i = - i \kappa \left( \zeta \sigma^m \bar\epsilon 
- \epsilon \sigma^m \bar\zeta \right) \partial_m \tilde\phi_i. 
\ee
Therefore, a condition $\tilde\phi_i(x) = {\rm constant}$ is 
invariant under supertransformations. 
\par
The explicit form of the relation between $\tilde C, \tilde\chi, 
\bar{\tilde\chi}, \cdots$ and $C, \chi, \bar\chi, \cdots$ is 
given by 
\ba
\tilde C \A = \A C' - i \kappa \zeta \chi' 
+ i \kappa \bar\zeta {\bar\chi}' 
+ {1 \over 2} i \kappa^2 \zeta^2 (M'+iN') 
- {1 \over 2} i \kappa^2 {\bar\zeta}^2 (M'-iN') \nonu
\A\A - \kappa^2 \zeta \sigma^m \bar\zeta v_m' 
- i \kappa^3 \zeta^2 \bar\zeta {\bar\lambda}' 
+ i \kappa^3 {\bar\zeta}^2 \zeta \lambda' 
+ {1 \over 2} \kappa^4 \zeta^2 {\bar\zeta}^2 D', \nonu
\tilde\chi \A = \A \chi' - \kappa \zeta (M'+iN') 
- i \kappa \sigma^m \bar\zeta v_m' 
+ 2 \kappa^2 \zeta \bar\zeta {\bar\lambda}' 
- \kappa^2 \lambda' {\bar\zeta}^2 
+ i \kappa^3 \zeta {\bar\zeta}^2 D', \nonu
\tilde M + i \tilde N \A = \A M' + i N' 
- 2 \kappa \bar\zeta {\bar\lambda}' 
- i \kappa^2 {\bar\zeta}^2 D', \nonu
\tilde v_m \A = \A v'_m - i \kappa \zeta \sigma_m {\bar\lambda}' 
+ i \kappa \lambda' \sigma_m \bar\zeta 
+ \kappa^2 \zeta \sigma_m \bar\zeta D', \nonu
\tilde\lambda + {1 \over 2} i \sigma^m \partial_m \bar{\tilde\chi} 
\A = \A \lambda' - i \kappa \zeta D', \nonu
\tilde D + {1 \over 2} \Box \tilde C 
\A = \A D', 
\label{tildefields}
\ea
where 
\ba
C' \A = \A C, \nonu
\chi' \A = \A \chi - \kappa \sigma^m \bar\zeta \partial_m C, \nonu
%
%
M' + i N' \A = \A M + i N 
+ i \kappa \partial_m \chi \sigma^m \bar\zeta 
- {1 \over 2} i \kappa^2 {\bar\zeta}^2 \Box C, \nonu
v'_m \A = \A v_m + {1 \over 2} \kappa \zeta \sigma^n 
\bar\sigma_m \partial_n \chi 
+ {1 \over 2} \kappa \partial_n \bar\chi \bar\sigma_m 
\sigma^n \bar\zeta 
- {1 \over 2} \kappa^2 \zeta \sigma^k \bar\sigma_m \sigma^l 
\bar\zeta \partial_k \partial_l C, \nonu
\lambda' 
\A = \A \lambda + {1 \over 2} i \sigma^m \partial_m \bar\chi 
- {1 \over 2} i \kappa \sigma^m \bar\zeta \partial_m (M-iN) 
+ {1 \over 2} \kappa \sigma^m {\bar\sigma}^n \zeta 
\partial_n v_m \nonu
\A\A - {1 \over 2} \kappa^2 \sigma^n \bar\zeta \zeta \sigma^m 
\partial_m \partial_n \bar\chi 
- {1 \over 4} \kappa^2 \Box \chi \zeta^2 
+ {1 \over 4} \kappa^3 \sigma^m \bar\zeta \zeta^2 
\partial_m \Box C, \nonu
D' \A = \A D + {1 \over 2} \Box C 
+ \kappa \zeta \sigma^n \partial_n \left( 
\bar\lambda + {1 \over 2} i {\bar\sigma}^m \partial_m \chi \right) 
- \kappa \bar\zeta {\bar\sigma}^n \partial_n \left( 
\lambda + {1 \over 2} i \sigma^m \partial_m \bar\chi \right) \nonu
\A\A + {1 \over 4} i \kappa^2 \zeta^2 \Box (M+iN) 
- {1 \over 4} i \kappa^2 {\bar\zeta}^2 \Box (M-iN) 
+ {1 \over 2} \kappa^2 \zeta \sigma^k {\bar\sigma}^m \sigma^l 
\bar\zeta \partial_k \partial_l v_m \nonu
\A\A - {1 \over 4} \kappa^3 \zeta^2 \partial_m \Box \chi 
\sigma^m \bar\zeta 
- {1 \over 4} \kappa^3 {\bar\zeta}^2 \zeta \sigma^m 
\partial_m \Box \bar\chi 
+ {1 \over 8} \kappa^4 {\bar\zeta}^2 \zeta^2 \Box^2 C. 
\ea
As in refs.\ \cite{IK}, \cite{UZ} it is possible to solve 
eq.\ (\ref{tildefields}) and express $C, \chi, \bar\chi, \cdots$ 
in terms of $\tilde C, \tilde\chi, \bar{\tilde\chi}, \cdots$ 
and $\zeta$, $\bar\zeta$. By imposing a SUSY and gauge invariant 
constraint on $\tilde\lambda$ the original 
fields $C, \chi, \bar\chi, \cdots$ 
become functions of $\tilde C$, $\tilde\chi$, 
$\bar{\tilde\chi}$, $\tilde M$, $\tilde N$, $\tilde v_m$, 
$\tilde D$ and $\zeta$, $\bar\zeta$. 
Substituting these expressions into an action one obtains 
an action of the N-G fields $\zeta$, $\bar\zeta$ interacting 
with other fields. Indeed, the couplings of $\zeta$, $\bar\zeta$ 
to $\tilde v_m$ were obtained in ref.\ \cite{IK}. 
Here, we are only interested in the sector 
which only depends on the N-G fields. 
\par
To eliminate other degrees of freedom than the N-G fields 
we impose SUSY invariant constraints 
\be
\tilde C = \tilde\chi = \tilde M = \tilde N = \tilde v_m 
= \tilde \lambda = 0, \qquad
\tilde D = {1 \over \kappa}. 
\label{constraint}
\ee
Solving these constraints we find that the original component 
fields $C$, $\chi$, $\bar\chi$, $\cdots$ can be 
expressed by the N-G fields $\zeta$, $\bar\zeta$. We find  
\ba
C \A = \A {1 \over 2} \kappa^3 \zeta^2 \bar\zeta^2, \nonu
\chi \A = \A -i \kappa^2 \zeta \bar\zeta^2 
+ \kappa \sigma^m \bar\zeta \partial_m C, \nonu
%
%
M+iN \A = \A -i \kappa \bar\zeta^2 
- i \kappa \partial_m \chi \sigma^m \bar\zeta 
+ {1 \over 2} i \kappa^2 \bar\zeta^2 \Box C, \nonu
%
%
v_m \A = \A \kappa \zeta \sigma_m \bar\zeta 
- {1 \over 2} \kappa \zeta \sigma^n \bar\sigma_m \partial_n \chi 
- {1 \over 2} \kappa \partial_n \bar\chi \bar\sigma_m 
\sigma^n \bar\zeta 
+ {1 \over 2} \kappa^2 \zeta \sigma^k \bar\sigma_m \sigma^l 
\bar\zeta \partial_k \partial_l C, \nonu
\lambda \A = \A i \zeta 
- {1 \over 2} i \sigma^m \partial_m \bar\chi 
+ {1 \over 2} i \kappa \sigma^m \bar\zeta \partial_m ( M-iN ) 
- {1 \over 2} \kappa \sigma^m \bar\sigma^n \zeta \partial_n v_m \nonu
\A\A + {1 \over 2} \kappa^2 \sigma^n \bar\zeta \zeta \sigma^m 
\partial_m \partial_n \bar\chi 
+ {1 \over 4} \kappa^2 \Box \chi \zeta^2 
- {1 \over 4} \kappa^3 \sigma^m \bar\zeta \zeta^2 
\partial_m \Box C, \nonu
%
%
D \A = \A {1 \over \kappa} - {1 \over 2} \Box C 
- \kappa \zeta \sigma^n \partial_n \left( \bar\lambda 
+ {1 \over 2} i \bar\sigma^m \partial_m \chi \right) 
+ \kappa \bar\zeta \bar\sigma^n \partial_n \left( \lambda 
+ {1 \over 2} i \sigma^m \partial_m \bar\chi \right) \nonu
\A\A - {1 \over 4} i \kappa^2 \zeta^2 \Box (M+iN) 
+ {1 \over 4} i \kappa^2 \bar\zeta^2 \Box (M-iN) 
- {1 \over 2} \kappa^2 \zeta \sigma^k \bar\sigma^m \sigma^l 
\bar\zeta \partial_k \partial_l v_m \nonu
\A\A + {1 \over 4} \kappa^3 \zeta^2 \partial_m \Box \chi 
\sigma^m \bar\zeta 
+ {1 \over 4} \kappa^3 \bar\zeta^2 \zeta \sigma^m 
\partial_m \Box \bar\chi 
- {1 \over 8} \kappa^4 \zeta^2 \bar\zeta^2 \Box^2 C. 
\label{relation}
\ea
The first equation gives $C$ in terms of $\zeta$, $\bar\zeta$. 
Substituting this into the second equation gives $\chi$ 
in terms of $\zeta$, $\bar\zeta$. 
By substituting these results into the third equation 
gives $M+iN$ in terms of $\zeta$, $\bar\zeta$, and so on.  
By the supertransformation of $\zeta$, $\bar\zeta$ in 
eq.\ (\ref{zetasupertrans}) these $C$, $\chi$, 
$\bar\chi$, $\cdots$ transform exactly as in 
eq.\ (\ref{supertrans}). 
The leading terms in the expansion of the fields $v_m$, $\lambda$, 
$\bar\lambda$ and $D$, which contain gauge invariant degrees of 
freedom, in $\kappa$ are 
\ba
v_m \A = \A \kappa \zeta \sigma_m \bar\zeta + \cdots, \nonu
\lambda \A = \A i \zeta 
- {1 \over 2} \kappa^2 \zeta 
\left( \zeta \sigma^m \partial_m \bar\zeta 
- \partial_m \zeta \sigma^m \bar\zeta \right) 
+ \kappa^2 \sigma^{mn} \zeta \partial_m 
\left( \zeta \sigma_n \bar\zeta\right) + \cdots, \nonu
%
%
D \A = \A {1 \over \kappa} 
+ i \kappa \left( \zeta \sigma^m \partial_m \bar\zeta 
- \partial_m \zeta \sigma^m \bar\zeta \right) + \cdots, 
\label{relation2}
\ea
where $\cdots$ are higher order terms in $\kappa$. 
In the four-component spinor notation the first equation becomes 
$v_m \sim \kappa \bar\zeta \gamma_m \gamma_5 \zeta + \cdots$, 
which is an axial vector. 
\par
%
%
Our discussion so far does not depend on a particular form 
of the action. We now consider a free action of a U(1) gauge 
supermultiplet with a Fayet-Iliopoulos $D$ term 
\be
S = {1 \over 4} \int d^4 x d^2 \theta \, W^\alpha W_\alpha 
+ {1 \over 4} \int d^4 x d^2 \bar\theta \, 
{\bar W}_{\dot\alpha} {\bar W}^{\dot\alpha} 
- {2 \over \kappa} \int d^4 x d^2 \theta d^2 \bar\theta \, V, 
\label{action1}
\ee
where 
\ba
W_\alpha \A = \A - {1 \over 4} {\bar D}_{\dot\beta} 
{\bar D}^{\dot\beta} D_\alpha V, \qquad
{\bar W}_{\dot\alpha} = - {1 \over 4} D^\beta D_\beta 
{\bar D}_{\dot\alpha} V, 
\nonu
D_\alpha \A = \A {\partial \over \partial \theta^\alpha} 
+ i (\sigma^m \bar\theta)_\alpha \partial_m, \qquad
\bar D_{\dot\alpha} 
= -{\partial \over \partial {\bar\theta}^{\dot\alpha}} 
- i (\theta \sigma^m)_{\dot\alpha} \partial_m. 
\ea
The last term proportional to $\kappa^{-1}$ is the 
Fayet-Iliopoulos $D$ term. In component fields we have 
\be
S = \int d^4x \left[ -{1 \over 4} v_{mn} v^{mn} 
- i \lambda \sigma^m \partial_m \bar\lambda 
+ {1 \over 2} D^2 - {1 \over \kappa} D \right]. 
\ee
The field equation for $D$ gives 
$D = {1 \over \kappa} \not= 0$ in accordance with 
eq.\ (\ref{relation2}), 
which shows that supersymmetry is spontaneously broken. 
\par
We substitute eq.\ (\ref{relation}) into the action 
(\ref{action1}) and obtain an action for the N-G fields 
$\zeta$, $\bar\zeta$. To do this it is more convenient to 
use a different form of the action equivalent to 
eq.\ (\ref{action1}) \cite{WB}
\be
S = \int d^4x d^2\theta d^2\bar\theta \, 
{\cal L}(x, \theta, \bar\theta), 
\ee
where 
\be
{\cal L} = -{1 \over 16} \left( {\bar D}^2 D^\alpha V D_\alpha V 
+ D^2 \bar D_{\dot\alpha} V \bar D^{\dot\alpha} V \right) 
- {2 \over \kappa} V. 
\ee
Changing the integration variables $(x, \theta, \bar\theta) 
\rightarrow (x', \theta', \bar\theta')$ by eq.\ (\ref{cov}) 
we obtain 
\ba
S \A = \A \int d^4x' d^2\theta' d^2\bar\theta' \, 
{\cal L}(x', \theta', \bar\theta') \nonu
\A = \A \int d^4x d^2\theta d^2\bar\theta 
J(x, \theta, \bar\theta) 
\tilde{\cal L}(x, \theta, \bar\theta), 
\label{action2}
\ea
where $J(x,\theta,\bar\theta)$ is the Jacobian for the 
change of variables and 
\be
\tilde{\cal L}(x, \theta, \bar\theta) 
= -{1 \over 16} \left( {\bar D'}{}^2 
{D'}^\alpha {\tilde V} D'_\alpha {\tilde V} 
+ {D'}{}^2 {\bar D}'_{\dot\alpha} {\tilde V} 
{\bar D}'{}^{\dot\alpha} {\tilde V} \right) 
- {2 \over \kappa} {\tilde V}. 
\ee
{}From eqs.\ (\ref{tildeexp}), (\ref{constraint}) we have 
\be
\tilde V = {1 \over 2\kappa} \theta^2 \bar\theta^2. 
\label{vtilde}
\ee
\par
In terms of the transformation matrix for the change of variables 
(\ref{cov}) 
\ba
M \A = \A {\partial(x', \theta', \bar\theta') 
\over \partial(x, \theta, \bar\theta)} \nonu
\A = \A \left( 
\begin{array}{ccc}
\delta_m^n - i \kappa \left( \theta \sigma^n \partial_m \bar\zeta 
- \partial_m \zeta \sigma^n \bar\theta \right) &
- \kappa \partial_m \zeta^\beta &
- \kappa \partial_m \bar\zeta^{\dot\beta} \\
- i \kappa (\sigma^n \bar\zeta)_\alpha & 
\delta_\alpha^\beta & 0 \\
- i \kappa (\zeta \sigma^n)_{\dot\alpha} & 0 & 
\delta_{\dot\alpha}^{\dot\beta} 
\end{array}
\right) 
\ea
the Jacobian and the transformation of derivatives are 
given by 
\be
J(x,\theta,\bar\theta) = {\rm sdet} M, \qquad
\left( 
\begin{array}{c}
{\partial \over \partial x'} \\ 
{\partial \over \partial \theta'} \\ 
{\partial \over \partial \bar\theta'} 
\end{array}
\right)
= M^{-1} 
\left( 
\begin{array}{c}
{\partial \over \partial x} \\ 
{\partial \over \partial \theta} \\ 
{\partial \over \partial \bar\theta} 
\end{array}
\right), 
\ee
where sdet is the superdeterminant. 
More explicitly, we obtain 
\ba
J \A = \A \det\left( V_m{}^n \right), \nonu
{\partial \over \partial x^{'m}} 
\A = \A V_m{}^n \left( {\partial \over \partial x^n} 
+ \kappa \partial_n \zeta^\beta 
{\partial \over \partial \theta^\beta} 
+ \kappa \partial_n {\bar\zeta}^{\dot\beta} 
{\partial \over \partial {\bar\theta}^{\dot\beta}} \right), \nonu
D'_\alpha 
\A = \A {\partial \over \partial \theta^\alpha} 
+ i (\sigma^n \bar\theta)_\alpha 
{\partial \over \partial x^{'n}}, \nonu
{\bar D}'_{\dot\alpha} 
\A = \A - {\partial \over \partial {\bar\theta}^{\dot\alpha}} 
- i (\theta \sigma^n)_{\dot\alpha} 
{\partial \over \partial x^{'n}}, 
\label{trans}
\ea
where 
\be
V_m{}^n = 
\delta_m^n 
- i \kappa \left( \theta \sigma^n \partial_m \bar\zeta 
- \partial_m \zeta \sigma^n \bar\theta \right) 
+ i \kappa^2 \left( \zeta \sigma^n \partial_m \bar\zeta 
- \partial_m \zeta \sigma^n \bar\zeta \right). 
\ee
Substituting eqs.\ (\ref{vtilde}), (\ref{trans}) into 
eq.\ (\ref{action2}) and integrating over $\theta$, $\bar\theta$ 
we obtain an action for the N-G fields 
\be
S = -{1 \over 2\kappa^2} \int d^4x \, \det \left[ \delta_m^n 
+ i \kappa^2 \left( \zeta \sigma^n \partial_m \bar\zeta 
- \partial_m \zeta \sigma^n \bar\zeta \right) \right]. 
\ee
This is exactly the V-A action. 
\par
Now we summarize the results as follows. 
All component fields of the vector gauge supermultiplet of 
linear SUSY are represented uniquely in terms of the N-G 
spinor field, and the V-A action of nonlinear SUSY is reproduced 
by just substituting the representations into the action 
of the vector gauge supermultiplet of linear SUSY. 
It is remarkable that the coefficients of all terms including 
the Fayet-Iliopoulos $D$ term in the linear SUSY action is 
determined uniquely by the SUSY (constraints). 
As for the axial vector nature of the U(1) gauge field 
we speculate that the adopted constraints may cut out implicitly 
the dyonic (electric and magnetic) aspect of the dynamics of the 
V-A action. All these phenomena are favorable to the SGM 
scenario \cite{KS1}.
%
%
\bigskip

The work of Y.T. is supported in part by the Grant-in-Aid from the  
Ministry of Education, Culture, Sports, Science and Technology, 
Japan, Priority Area (\#707) ``Supersymmetry and Unified Theory 
of Elementary Particles''. 
The work of M.T. is supported in part by the High-Tech research 
program of Saitama Institute of Technology. 

%
%
\newcommand{\NP}[1]{{\it Nucl.\ Phys.\ }{\bf #1}}
\newcommand{\PL}[1]{{\it Phys.\ Lett.\ }{\bf #1}}
\newcommand{\CMP}[1]{{\it Commun.\ Math.\ Phys.\ }{\bf #1}}
\newcommand{\MPL}[1]{{\it Mod.\ Phys.\ Lett.\ }{\bf #1}}
\newcommand{\IJMP}[1]{{\it Int.\ J. Mod.\ Phys.\ }{\bf #1}}
\newcommand{\PR}[1]{{\it Phys.\ Rev.\ }{\bf #1}}
\newcommand{\PRL}[1]{{\it Phys.\ Rev.\ Lett.\ }{\bf #1}}
\newcommand{\PTP}[1]{{\it Prog.\ Theor.\ Phys.\ }{\bf #1}}
\newcommand{\PTPS}[1]{{\it Prog.\ Theor.\ Phys.\ Suppl.\ }{\bf #1}}
\newcommand{\AP}[1]{{\it Ann.\ Phys.\ }{\bf #1}}
\newcommand{\ATMP}[1]{{\it Adv.\ Theor.\ Math.\ Phys.\ }{\bf #1}}
\end{document}